\documentclass[final, authoryear, 5p, times, twocolumn]{elsarticle}

\usepackage{graphicx}
\usepackage{amssymb}
\usepackage{amsmath}
\usepackage[english]{babel}
\usepackage{lineno}
\usepackage{threeparttable}
\usepackage{tabularx}
\usepackage{color}
\usepackage[table]{xcolor}
\usepackage{mathtools, cuted}
\usepackage{multirow}

\newcolumntype{L}[1]{>{\raggedright\arraybackslash}p{#1}}
\newcolumntype{C}[1]{>{\centering\arraybackslash}p{#1}}
\newcolumntype{R}[1]{>{\raggedleft\arraybackslash}p{#1}}

\journal{arXiv.org}

\begin{document}

\begin{frontmatter}

\title{On the algebraic definition of total rotation in RSA}

\author[scstc]{Marco ~Bontempi\corref{cor1}}

\cortext[cor1]{Corresponding author. E-mail address: marco.bontempi@ior.it; Tel.: +39 051 636 6852; fax: +39 051 583789}

\address[scstc]{Struttura Complessa di Scienze e Tecnologie Chirurgiche, IRCCS Istituto Ortopedico Rizzoli, via di Barbiano 1/10, I-40136, Bologna (BO), Italy}



\date{\today}

\begin{abstract}
Total rotation is a quantity that has been used for years in RSA.
However, its definition has no mathematical sense, since the Euler angles do not form a vector space, since angles cannot define a multiplication group.
With this work I tried to give a mathematical definition of the total rotation connecting the Euler description of the rotations with the helical axis. 
The approximation for small angles was used to connect Euler's angles and helical angle.
With this approximation Euler angles acquire the properties of a vector space and it is possible to justify the meaning of this parameter.
Validation test showed that total rotation has an approximation error between 5\% and 7\% for angles in the range $\left[-\frac{\pi}{6}, \frac{\pi}{6} \right]$. 
Since usually RSA uses smaller angle ranges, the approximation is perfectly suitable for use in RSA.
\end{abstract}


\end{frontmatter}

\section{Introduction} \label{sec:intro}
In Roentgen Stereophotogrammetric Analysis (RSA) it is essential to measure the ``migration'' \citep{Selvik1989, Valstar2005} to determine the stability of a prosthetic implant, or in general to evaluate the micro-movement of one object compared to another.
Migration is expressed in terms of "total translation" and "rotational rotation" \citep{Ryd1986, Selvik1989}.
Rotations were usually described in terms of Euler decomposition around three cardinal axes properly oriented.

While the "total translation" is well defined by the norm of the displacement vector between  initial and final position, the "total rotation" does not have a strict mathematical definition, but only an operative expression.
The requested quantity should to define the rotation of one object from an in initial position to its final position using a single value as the total translation do.
Operationally, in analogy with total translation, total rotation was defined by applying the norm to the Euler angles obtained from data processing:
\begin{equation}
	\label{eq:rotation}
	\theta_{T} = \sqrt{\theta_{x}^{2} + \theta_{y}^{2} + \theta_{z}^{2}}
\end{equation}
where $\theta_{T}$ is the total angle, and $\theta_{x, y, z}$ are the rotation angles derived by the Euler decomposition of the object positions.
From the algebra point of view, this definition is wrong.
The reason is because Euler's angles cannot represent a vector space \cite{Meckes2018}.
Particularly the angle's multiplication cannot be defined and angles are not a group for this operation.
This is due to the periodic structure of angles and also by the fact that trigonometric functions are not bijective.

An other way to describe rotations in space is the helical angle/axis, or screw angle/axis \citep{Taylor1994, MillanVaquero2016}.
It is defined as the rotation angle around an axis.
In this way every rotation can be defined using a single scalar value: the helical angle (screw angle).
This seems to be a good candidate to describe rotations using a single scalar. 
But its representation is totally different from the Euler representation.

\cite{Selvik1989} discussed the connection between kinematic description of small movements and $\theta$, providing an implicit definition of total rotation. 
The aim of this report is to review Selvik work from a the algebra point of view to find a connection between Euler angles and helical angle and to adjust the formulation of total angle used in RSA in terms of these quantities.

\section{Materials and Methods} \label{sec:mm}
\subsection{Mathematical derivation of total rotations}
Euler decomposition can be applied to orthogonal matrices that represent transformations between to reference systems.
Let $M \in \mathbb{R}^{3 x 3}$ be a transformation matrix between to reference systems $R_{1}$ and $R_{2}$ in the Euclidean space $\mathbb{R}^{3}$.
Thus, three matrices $E_{x}$, $E_{y}$, $E_{z}$ exist that represent elementary transformations around cardinal axes X, Y, and Z.
These three matrices can be expressed as:
\begin{equation}
	\label{eq:ematrices}
	\left\{ \begin{array}{l}
		E_{x}(\alpha) = \left( \begin{array}{ccc}
			1 & 0 & 0 \\
			0 & \cos \alpha & -\sin \alpha \\
			0 & \sin \alpha & \cos \alpha
		\end{array} \right) \\ \\
		E_{y}(\beta) = \left( \begin{array}{ccc}
			\cos \beta & 0 & \sin \beta \\
			0 & 1 & 0 \\
			-\sin \beta & 0 & \cos \beta
		\end{array} \right) \\ \\
		E_{z}(\gamma) = \left( \begin{array}{ccc}
			\cos \gamma & -\sin \gamma & 0\\
			\sin \gamma & \cos \gamma & 0 \\
			0 & 0 & 1
		\end{array} \right)
	\end{array} \right.
\end{equation}
where $\alpha$, $\beta$ and $\gamma$ are called Euler's angles.
The transformation $M$ can be expressed as a combination of these three matrices.
All possible combinations are:
\begin{equation}
	\label{eq:comb}
	M = \left\{ \begin{array}{l}
		E_{x}E_{y}E_{z} \\
		E_{y}E_{z}E_{x} \\
		E_{z}E_{x}E_{y} \\
		E_{x}E_{z}E_{y} \\
		E_{z}E_{y}E_{x} \\
		E_{y}E_{x}E_{z} \\
		\\
		E_{z}E_{x}E_{z} \\
		E_{x}E_{y}E_{x} \\
		E_{y}E_{z}E_{y} \\
		E_{z}E_{y}E_{z} \\
		E_{x}E_{z}E_{x} \\
		E_{y}E_{x}E_{y}
	\end{array} \right.
\end{equation}
The first six combinations are called ``Tait-Bryan combinations'', while the second six are called ``Proper Euler combinations''.
Each combination has its Euler's angles $(\alpha, \beta, \gamma)$, called Tait-Bryan angles, and Proper Euler angles.

The helical representation of M is defined by an angle ($\xi$) and an axis, described by the components of the axis direction unit vector $\hat{u} = (u_{x}, u_{y}, u_{z})$.
This can be expressed as (Rodrigues'formula):
\begin{equation}
	\label{eq:hmatrix}
	M = \left( \cos \xi \right) \mathbb{I} + \left( \sin \xi \right) [u]_{\times} + \left( 1 - \cos \xi \right) \left( u \otimes u \right)
\end{equation}
where $\mathbb{I}$ is the identity matrix, $\left( u \otimes u \right)$ is the outer product:
\begin{equation}
	\label{eq:uoumatrix}
	\left( u \otimes u \right) = \left(  \begin{array}{c}
u_{x} \\ u_{y} \\ u_{z}
	\end{array} \right) \left( \begin{array}{ccc}
		u_{x} & u_{y} & u_{z} 
	\end{array} \right)
\end{equation}
and $[u]_{\times}$ is the cross product matrix of u:
\begin{equation}
	\label{eq:uxmatrix}
	[u]_{\times} = \left(  \begin{array}{ccc}
		0 & -u_{z} & u_{y} \\
		u_{z} & 0 & -u_{x} \\
		-u_{y} & u_{x} & 0 
	\end{array} \right)
\end{equation}
The helical angle can be derived by manipulating equation \ref{eq:hmatrix} and the result is:
\begin{equation}
	\label{eq:hangle}
	\xi = \arccos \frac{tr(M) - 1}{2}
\end{equation}
where $tr(M)$ is the trace of $M$.

It is possible to combine the Euler representation and the helical representation and extrapolate the relation between them.

The two set of possible combinations share two common helical angle representations: 
\begin{equation}
	\label{eq:cosxi}
	\cos \xi = \left\{ 
	\begin{array}{l}
		\frac{\cos \alpha \cos \beta + \sin \alpha \sin \beta \sin \gamma + \cos \alpha \cos \gamma + \cos \beta \cos \gamma - 1}{2}\\
		\\
		\frac{\cos \alpha \cos \beta \cos \gamma - \sin \alpha \sin \gamma + \cos \alpha \cos \gamma - \sin \alpha \cos \beta \sin \gamma + \cos \beta - 1}{2}
		\end{array} \right.
\end{equation}
These equations are the general form the helical angles as a function of the Euler's angles according to the matrix combination.

The angle $\xi$ is the most suitable candidate to represent the total rotation.
It summarizes, in fact, the Euler rotations into one single angular value.
But its form is complex and the $\arccos$ function requires great care to be handled properly.

As suggested by \cite{Selvik1989}, static RSA was designed to evaluate micro-motions between objects.
Thus, it is more practical to consider small angles approximation.
This allow to modify the expression in equations \ref{eq:cosxi}, because ``cos'' and ``sin'' functions can be approximated using Taylor's formulas:
\begin{equation}
	\label{eq:sincos}
	\left\{ \begin{array}{l}
		\cos \theta = 1 - \frac{\theta^{2}}{2} + o(\theta^{3}) \\ \\
		\sin \theta = \theta + o(\theta^{2})
	\end{array} \right.
\end{equation}
By combining equations \ref{eq:cosxi} and \ref{eq:sincos} the result is:
\begin{equation}
	\label{eq:Thabg}
	\xi \approx \sqrt{\alpha^{2} + \beta^{2} + \gamma^{2}}
\end{equation}
which is the same formulation of the total rotation expressed in equation \ref{eq:rotation}.
Thus, for small angles, total rotation, as used in RSA, is the helical angle.
Moreover, this representation is common to all possible combinations (see table \ref{tab:helicals}).
\begin{table*}[h]
	\centering 
	\begin{tabular}{c|c|c}
		\multirow{2}{*}{combination} & \multirow{2}{*}{exact helical angle ($\cos \xi$)} & approximated helical \\
		            &                                                   & angle ($\xi$) \\
		\hline
 		$E_{x}E_{y}E_{z}$ &&\\
		$E_{y}E_{z}E_{x}$ &&\\
		$E_{z}E_{x}E_{y}$ &\multirow{2}{*}{$\frac{\cos \alpha \cos \beta + \sin \alpha \sin \beta \sin \gamma + \cos \alpha \cos \gamma + \cos \beta \cos \gamma - 1}{2}$}&\\
		$E_{x}E_{z}E_{y}$ &&\\
		$E_{z}E_{y}E_{x}$ &&\\
		$E_{y}E_{x}E_{z}$ & & \multirow{2}{*}{$\sqrt{\alpha^{2} + \beta^{2} + \gamma^{2}}$}\\
		\cline{1-2}
		$E_{z}E_{x}E_{z}$ &&\\
		$E_{x}E_{y}E_{x}$ &&\\
		$E_{y}E_{z}E_{y}$ &\multirow{2}{*}{$\frac{\cos \alpha \cos \beta \cos \gamma - \sin \alpha \sin \gamma + \cos \alpha \cos \gamma - \sin \alpha \cos \beta \sin \gamma + \cos \beta - 1}{2}$} &\\
		$E_{z}E_{y}E_{z}$ &&\\
		$E_{x}E_{z}E_{x}$ &&\\
		$E_{y}E_{x}E_{y}$ &&\\
		\hline
	\end{tabular}
	\caption{\label{tab:helicals}Summary of exact solution of helical angle form and approximations of all possible Euler matrices combinations.}
\end{table*}

\subsection{Approximation comparison}
Equations listed in table \ref{tab:helicals} are two aspects of the same thing.
The exact solution is valid for every value of $\alpha$, $\beta$ and $\gamma$, while the approximated solution can be considered valid only within a limited set of values of the three angles.

To make the definition of total rotation usable, it is necessary to compare its value with the exact helical angle. 
In this way it is possible to understand the range of angles that allow to use its definition.
To analyze the differences between exact helical angle and total rotation (approximated helical angle) a set of test were performed.
Test were performed by implementing a set of different angle ranges and applying them to the exact and approximated equations of the helical axis.
Test were implemented using MATLAB\textregistered (R2018a, MathWorks, Natick, MA, USA).

The comparison was calculated as the value of the helical angle over and below the principal bisector of plane divided by the corresponding value of the total rotation.
\begin{equation}
	\label{eq:ratio}
	\Delta(\xi_{ha}^{\pm}, \xi_{tr}^{\pm}) = \frac{\xi_{ha}^{\pm}}{\xi_{tr}^{\pm}}
\end{equation}

\section{Results} \label{sec:results}
\begin{figure}[h]
	\begin{center}
		\includegraphics[width=8.6cm]{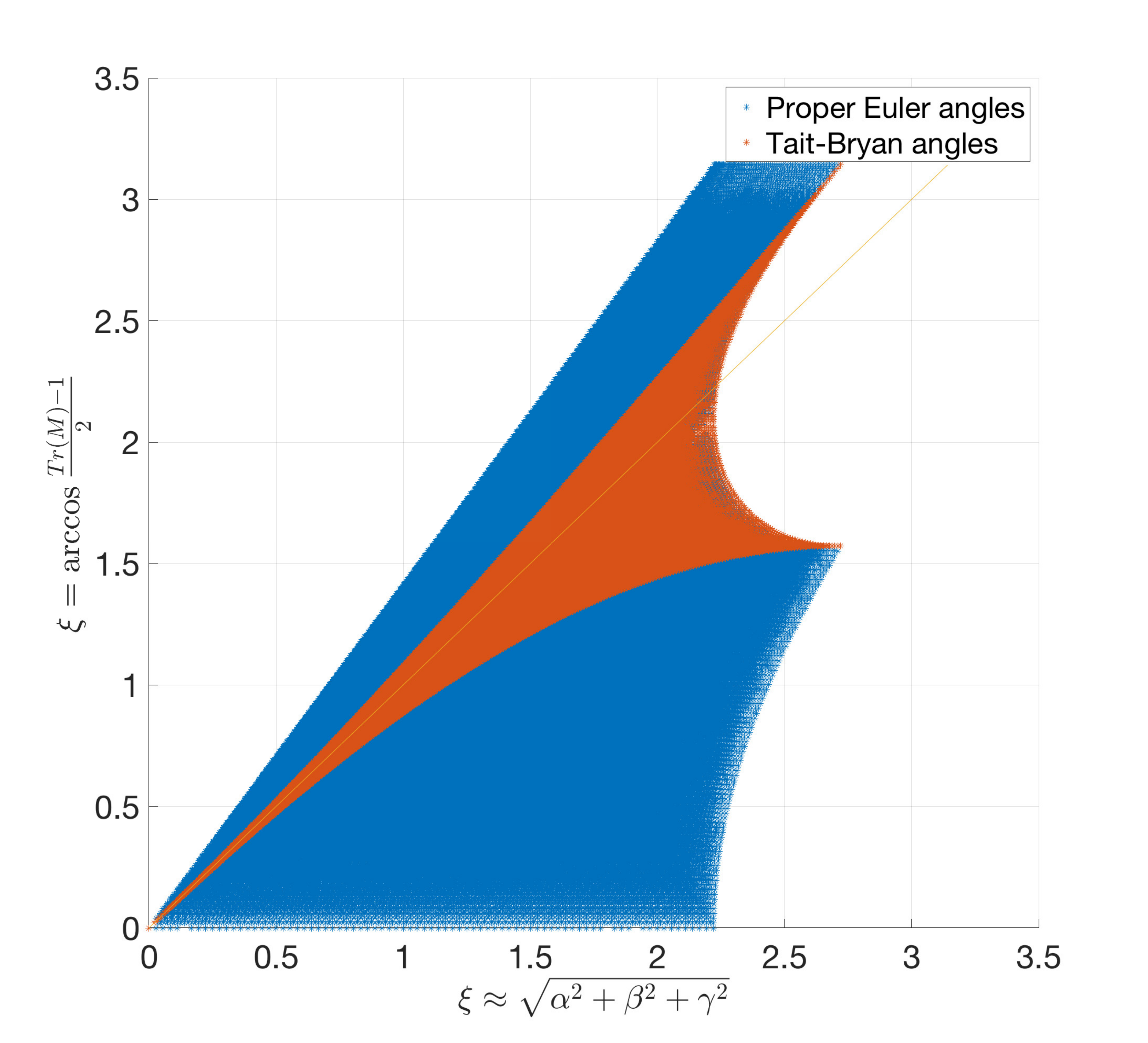}
		\caption{\label{fig:comp2pi} Comparison of exact helical angle solution and approximated values calculated according to angles $\{\alpha, \beta, \gamma\} \in [-\frac{\pi}{2}, \frac{\pi}{2}]$. The main bisector rest represent the exact correlation between the two representations.}
	\end{center}
\end{figure}

\begin{figure}[h]
	\begin{center}
		\includegraphics[width=8.6cm]{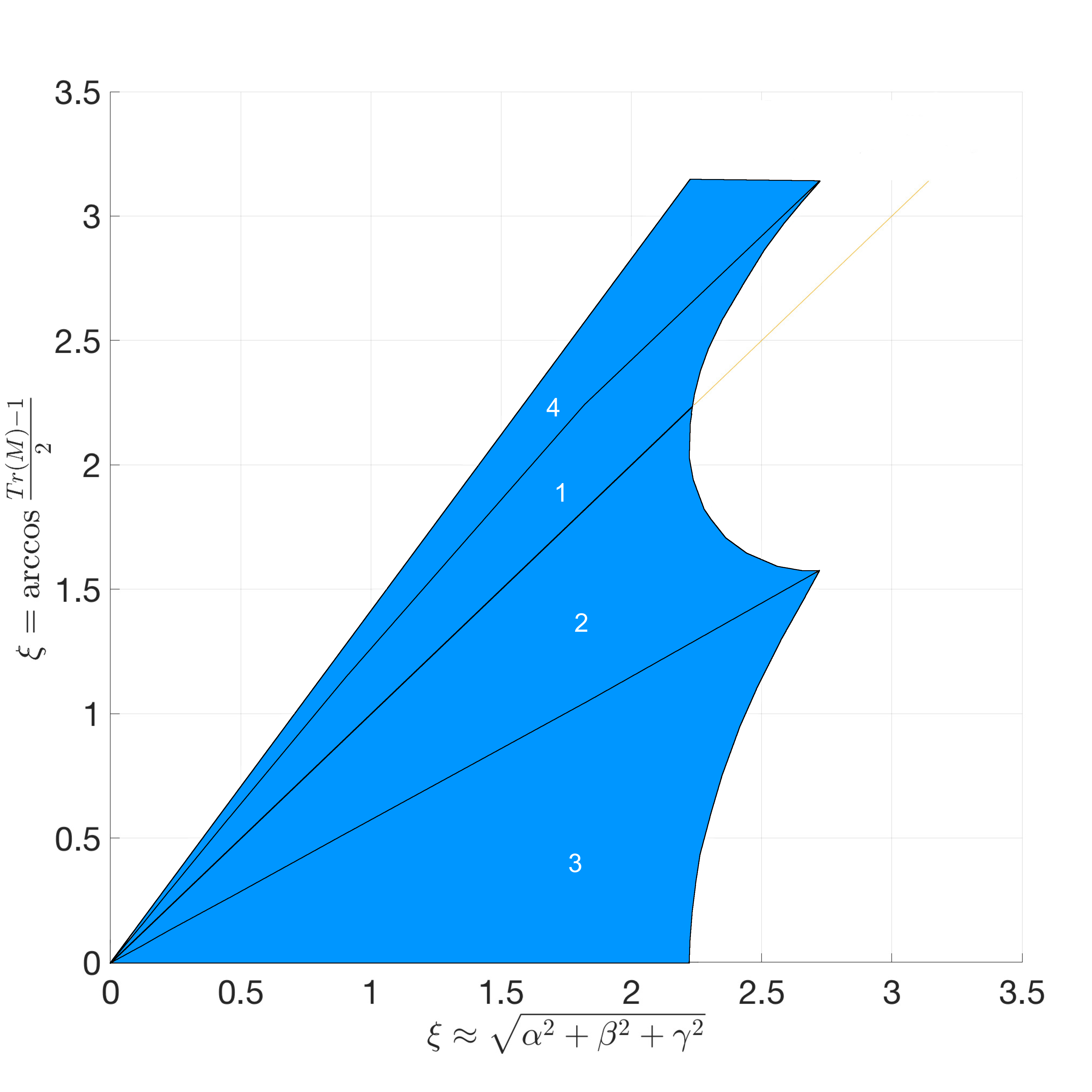}
		\caption{Zones of different Euler combination angles that affect the correlation between Proper Euler combination helical angle and total rotation approximation. Zone 1 identify angles in the range $[0, \frac{\pi}{2}]$. Zone 2 refers to angle range $[-\frac{\pi}{2}, 0]$. Zone 3 was generated by angle pattern made of $(\pm\alpha, \beta, \mp\alpha)$, and zone 4 contains all other combinations.}
		\label{fig:zones}
	\end{center}
\end{figure}

Figure \ref{fig:comp2pi} shows the comparison between exact and approximated helical angles solutions.
On the horizontal axis is the total rotation and the vertical axis shows the exact helical angle.
The main bisector line of the plane is a reference for estimating the differences between the two representations.
The figure show significant differences between the results obtained with the "Tait-Bryan" and "Proper Euler" combinations.
Particularly, Tiat-Bryan combinations are closer to the total rotation. 
On the contrary, Proper Euler decompositions have a wider range of values and highlights a not good approximation.
This great variability is due to the formulation of helical angle for Proper Euler combinations. 
The combination of angles strongly affect the shape of the helical angle and different zones can be identified, as shown in figure \ref{fig:zones}.

Concerning the accuracy of total rotation and helical angle, the results are listed in table \ref{tab:diff}.
\begin{table}[h]
	\centering 
	\begin{tabular}{c|c|c|c|c}
		angle & \multicolumn{2}{c|}{Tiat-Bryan} & \multicolumn{2}{c}{Proper Euler} \\  
		range & $\Delta(\xi_{ha}^{-}, \xi_{tr}^{-})$ & $\Delta(\xi_{ha}^{+}, \xi_{tr}^{+})$ & $\Delta(\xi_{ha}^{-}, \xi_{tr}^{-})$ & $\Delta(\xi_{ha}^{+}, \xi_{tr}^{+})$ \\
		      & (\%) & (\%) & (\%) & (\%) \\
		\hline
		&&&&\\
   		$[-\frac{\pi}{2},  \frac{\pi}{2} ]$ & 57.7 & 115.5 & 57.7 & 115.5 \\ &&&&\\
   		$[-\frac{\pi}{3},  \frac{\pi}{6} ]$ & 75.5 & 112.3 & 57.7 & 123.8 \\ &&&&\\
   		$[-\frac{\pi}{6},  \frac{\pi}{6} ]$ & 89.6 & 107.3 & 57.7 & 127.9 \\ &&&&\\
   		$[-\frac{\pi}{12}, \frac{\pi}{12}]$ & 95.3 & 104.0 & 57.7 & 128.8 \\ &&&&\\
   		$[-\frac{\pi}{24}, \frac{\pi}{24}]$ & 97.8 & 102.1 & 57.7 & 129.0 \\ &&&&\\
   		$[-\frac{\pi}{48}, \frac{\pi}{48}]$ & 98.9 & 101.1 & 57.7 & 129.0 \\ &&&&\\
		\hline
	\end{tabular} 
	\caption{\label{tab:diff} list of differences between heliacal angle and total rotation. Proper Euler combinations considered only points in regions 1 and 2. Values closer to 100\% denote better approximation.}
\end{table}

\section{Discussion} \label{sec:discussion}
The approximation of small angles allowed to connect the Euler angles to the helical angle and the total rotation is the bridge between the two representations.

Euler matrices can be combined in 12 different ways and each combination has its own result.
Fortunately, all combinations can be grouped into 2 categories: Tiat-Bryan combination and Proper Euler combination.
The first category combines the 3 matrices without repeating the rotations around the cardinal axes. 
It is widely used in aeronautics and biomechanics.
The RSA itself uses this way to combine the Euler matrices. 
The second category involves the repetition of rotations around one of the cardinal axes. 
It is never used in biomechanics and is never considered in RSA.

The two sets of combinations have two different ways of representing helical angles as a function of Euler's angles.
Moving to the approximation for small angles, both helical angle formulations give the same result, shown in table \ref{tab:helicals}.
The total rotation is exactly this result.
 
The result obtained was then validated to compare accuracy and applicability of the approximation for small angles with the exact solution.
The result of the comparison is shown in figure \ref{fig:comp2pi}. 
The plot was obtained by calculating the helical axis defined in table \ref{tab:helicals} considering all possible combinations of Euler angles in the range $\left[ -\frac{\pi}{2}, \frac{\pi}{2} \right]$.

Figure \ref{fig:comp2pi} shows a great difference in the behavior of formulas in the angular range considered.
Particularly, the Tait-Bryan combinations show a much more regular and compact behavior around the main bisector of the plane, which represents the perfect correspondence between helical angle and total rotation.
On the contrary, the Proper Euler combinations show a very variable trend.
To understand this spread of results, Proper Euler angles were analyzed as a function of Euler angles ranges.
This analysis provided four zones (figure \ref{fig:zones}) that determines different responses of helical angles and total rotations.
In this category of combinations, therefore, the approximation for small angles is applicable only for some particular angle combinations.
Specifically, the combinations $\left[ -\frac{\pi}{2}, 0 \right]$ and $\left[ 0, \frac{\pi}{2} \right]$ (zones 1 and 2 in figure \ref{fig:zones}) give acceptable results, comparable with Tait-Bryan combinations.

A deeper analysis of the relationships between helical angle and total rotation showed that Tiat-Bryan combinations also suffer the same problem of dispersion of results, but in angular ranges $\left[ -\pi, -\frac{\pi}{2} \right]$ and $\left[ \frac{\pi}{2}, \pi \right]$ that are not usually used.

These important discrepancies were due to the nature of the angles and the fact that, as said, Euler's angles do not form a vector space.

Table 2 helps to define the validity range of the approximation for small angles. 
It is clear that the smaller the angle range the better the approximation. 
Particularly, for angles between $\left[ -\frac{\pi}{6}, \frac{\pi}{6} \right]$ the difference is less than 7\% for Tait-Bryan combinations that are those used in RSA.
Generally this angular range is much greater than the typical angles evaluated in RSA which are usually within the range $\left[ \frac{\pi}{36}, \frac{\pi}{36} \right]$ \citep{Selvik1989, Valstar2005}, where the difference drops to less than 1\%.
In this angular range, trigonometric functions become linear and it is possible to define a homomorphism between $(\alpha, \beta, \gamma)$ and $\mathbb{R}^{3}$ which is a vector space.
Thus, Euler angles acquire, locally, the properties of a vector space and it is possible to define a norm and the equation \ref{eq:rotation} is valid.

This was also considered by \cite{Selvik1989} which applied the small angle approximation to write the kinematic equation to describe micro movements.
Selvik's equations can be reformulated in terms of approximated Rodrigues'formula (equation \ref{eq:hmatrix}) and the equation \ref{eq:Thabg} can be easily derived.
 
Therefore, for the typical angular ranges of RSA the approximation for small angles of the helical angle calculated with Euler angles can be considered valid and the parameter "total rotation" acquires a rigorous mathematical meaning.

\section{Conclusion} \label{sec:conclusion}
This work tried to give a definition that would justify the total rotation that synthesizes the rotational migration of rigid bodies studied in RSA.
Moreover, the result connects the total rotation with the helical angle, the Euler's angles and the algebra of vector spaces, providing a wider and more precise picture of the relationships between these important kinematic variables.

\section*{Acknowledgement}
I would like to acknowledge Bart L. Kaptein for his assistance and the nice conversation we had on this topic.

\section*{Conflict of interest statement}
The author declares that they have no known competing financial interests or personal relationships that could have appeared to influence the work reported in this paper.

\section*{Funding}
This research did not receive any specific grant from funding agencies in the public, commercial, or not-for-profit sectors.


\end{document}